\documentstyle[aps,pre,multicol,epsf,tabularx]{revtex}
\def\be{\begin{equation}}
\def\ee{\end{equation}}
\def\bea{\begin{eqnarray}}
\def\eea{\end{eqnarray}}
\def\ba{\begin{array}}
\def\ea{\end{array}}
\def\figspace{\vspace{0.4cm}}

\begin{document}
\draft

\title{Phase Transitions and Oscillations 
	in a Lattice Prey-Predator Model}
\vspace {1truecm}
\author{Tibor Antal and Michel Droz}
\address{ D\'epartement de Physique Th\'eorique, Universit\'e de
	  Gen\`eve, CH 1211 Gen\`eve 4, Switzerland.}

\maketitle

\begin{abstract}
A coarse grained description of a two-dimensional prey-predator system
is given in terms of a 3-state lattice model containing two control parameters:
the spreading rates of preys and predators.
The properties of the model are investigated by
dynamical mean-field approximations and extensive numerical simulations.
It is shown that the stationary state phase diagram is divided into two phases:
a pure prey phase and a coexistence phase of preys and predators
in which temporal and spatial oscillations can be present.
The different type of phase transitions occuring at the boundary of the prey 
absorbing phase, as well as the crossover phenomena occuring between 
the oscillatory and non-oscillatory domains of the coexistence phase 
are studied.
The importance of finite size effects are discussed  and scaling relations 
between different quantities are established.
Finally, physical arguments, based on the spatial structure of the model,
are given to explain the underlying mechanism leading to oscillations.
\end{abstract}


\begin{multicols}{2}
\narrowtext

\section{Introduction}
\label{sec:intro}

The dynamics of interacting species has attracted a lot of attention 
since the pioneering works of Lotka~\cite{lotka}  and 
Volterra~\cite{volterra}. In their independent studies, they showed 
that simple prey-predator models may exhibit limit cycles during 
which the populations of both species have periodic oscillations in time. 
However, this behavior depends strongly on the initial state, and it
is not robust to the addition of more general 
non-linearities  or to the presence of more  than two interacting 
species~\cite{montrol}. In many cases the system reaches a simple 
steady-state.

A better understanding of the properties of such oscillations is clearly
desirable, as such  population cycles are often observed in ecological 
systems and the  underlying causes remain a long-standing open 
question~\cite{Nature}. 
One of the best documented example concerns the Canadian lynx population. 
This population was monitored for more than hundred years (starting in 1820) 
from different regions of Canada. It was observed that the population 
oscillates with a period of approximately 10 years and that this 
synchronization was spatially extended over areas of several millions 
of square kilometers~\cite{lynx}.
Several attempts were made to explain these facts 
(climatic effects, relations with the food-web, influence if the solar cycle) 
without success. More recently, Blasius et al.~\cite{Nature} introduced a 
deterministic three level vertical food-chain model. The three coupled 
nonlinear differential equations defining the model contain eight free 
parameters and two unknown nonlinear functions. 
The authors showed that an ad-hoc choice of the free parameters and  
nonlinear functions explains the experimental data for the Canadian lynx.

In such mean-field type models, it is assumed that the populations 
evolve homogeneously, which is obviously an oversimplification. 
An important question consists in understanding
the role played by the local environment on the dynamics~\cite{may}. 
There are many examples in equilibrium and nonequilibrium statistical 
physics showing that, in low enough dimensions, the local aspects 
(fluctuations) play a crucial role
and have some dramatic effects on the dynamics of the system.
Accordingly, a lot of activities have been devoted during 
the past years to the study of extended prey-predator 
models. The simplest spatial generalization are the so called two
patches models, where the species follow the conventional
prey predator rules within each patches, and can migrate from one
patch to the other~\cite{jansen}.
Other works have found that the introduction of stochastic dynamics
plays an important role~\cite{bradshaw},
as well as the use of discreet variables, which prevent the 
population to become vanishingly small.

These ingredients are included in the so called individual based
lattice models, for which each lattice site can be empty or
occupied by one \cite{{tome},{boccara},{provata},{tainaka},{pekalski}} 
individual of a given species or two \cite{{lipowski},{lipowska}} 
individuals belonging to different species.
It was recognized that these models
give a better description of the oscillatory behavior
than the usual Lotka-Volterra (L-V) equations.
Indeed, the oscillations in these lattice models are stable against small 
perturbations of the prey and predator densities,
and they do not depend on the initial state. 
It was also found (in two dimensional systems)
that the amplitude of the oscillations of global quantities decreases 
with increasing system size, while the oscillations persist on local level. 
It was  argued that coherent periodic oscillations are absent in large systems
(although, \cite{tome} do not discard this possibility).
In \cite{lipowski} Lipowski et al.\ state that this is only possible above 
a spatial dimension of 3. 
In \cite{provata} Provata et al.\ emphasize that the frequency
of the oscillations are stabilized by the lattice structure
and that it depends on the lattice geometry.
In some papers, the stationary phase diagram was also derived
\cite{{tome},{lipowska}},
and different phases were observed as a function of the model parameters,
such as an empty phase, a pure prey phase,
and an oscillatory region of coexisting preys and predators.
In \cite{tome}, a coexistence region without oscillations and
a domain of the control parameter space for which the stationary
states depend strongly upon the initial condition, were found.

However, in all the above works no systematic finite size studies 
have been performed, allowing to draw firm conclusions on the phase 
diagram of the models as a function of their sizes.
It is known~\cite{chopard}, that in ecological problems
the fact that a system has a finite size is more relevant
than in most of the cases encountered in statistical physics,
for which one concentrates on the thermodynamic limit.
Particularly, the size dependence of the amplitude of the oscillations,
as well as a detailed description of the 
critical behavior near the phase transitions have not been investigated. 
Another relevant question is how much the stationary phase diagrams of 
these prey-predator models have some generic properties or how 
much they depend upon the details of the models.

The goal of this paper is to study a simple models of prey-predators 
on a two-dimensional lattice for which  some of the above questions 
could be answered. 
Our model is based on a coarse-grained description  in the sense that 
a given cell models a rather large part of a territory and thus can 
contain many preys or predators. Moreover, predators cannot leave 
without preys in a given cell. 
Those are the main differences between our model and  Satulovsky 
and Tom\'e (ST) model ~\cite{tome}. Nevertheless, it turns out 
that the stationary state phase diagram of the two models are quite different.

Our model is defined in Sec.~\ref{sec:model}. Although governed by 
only two control parameters, this model exhibits a rich phase diagram.
Two different phases are observed: a pure prey phase,
and a coexistence phase of preys and predators
in which an oscillatory and a non-oscillatory region can be distinguished.
In some limiting cases the model can be mapped onto another well known
nonequilibrium model: the {\sl contact process} (CP)~\cite{contact}.
In Sec.~\ref{sec:mean-field} the properties of our model are analyzed 
in dynamical one and two-points mean-field approximations and no 
undamped oscillatory behavior is found. In Sec.~\ref{sec:mc}, extensive 
Monte-Carlo simulations are performed. 
It is shown that, as a function of the values of the control 
parameters,  two types of continuous nonequilibrium phase transitions 
towards a prey absorbing state are present. 
The system size dependence of the amplitude of the oscillations is 
studied and several scaling relations between the amplitude of the 
oscillations and the correlation length are obtained. 
In  Sec.~\ref{sec:discuss} an underlying mechanism responsible for the 
spatial oscillations is proposed, which leads to a qualitative  
explanation of the properties of the phase diagram.
In particular, we show that the spatially extended aspect of 
the problem is crucial to have an oscillatory region.
Finally, conclusions  are drawn in Sec.~\ref{sec:conc}.

\section{The model}
\label{sec:model}

Our system models preys and predators 
living together in a two dimensional territory.
This territory is divided into square cells,
and each of them can contain several preys and predators.
In this coarse-grained description in which each cell represents
a rather large territory, one can assume that each cell containing
some predator will also contain some preys.
Hence, a three state representation is made.
Each cell of the two-dimensional square lattice
(of size $L \times L$, with periodic boundary condition), 
labeled by the index $i$,  
can be at time $t$, in one of the three following states:
$\sigma_i = 0, 1, 2$.
A cell in state 0, 1 or 2 corresponds respectively to a cell which is empty,
occupied by preys or simultaneously occupied by preys and predators.
The dynamics of the system is defined as a continuous time 
Markov process.
The transition rates for site $i$ are
\begin{itemize}
\item[i)]
$0 \to 1$ at rate $\lambda_a (n_{i,1}+n_{i,2})/4$,
\item[ii)]
$1 \to 2$ at rate $\lambda_b (n_{i,2})/4$,
\item[iii)]
$2 \to 0$ at rate 1,
\end{itemize}
where $n_{i,\sigma}$ denotes the number of nearest neighbor sites 
of $i$ which are in the state $\sigma$. 4 is the coordination number of this 
two dimensional square lattice. 

The first two processes model the spreading of preys and predators.
The two control parameters, $\lambda_a$ and $\lambda_b$,
characterize a particular prey-predator system.
The reason for considering the sum,
$n_{i,1}+n_{i,2}$, in the first rule is simply that 
all the neighboring cells of $i$ containing some prey,
(hence $\sigma_i=1$ or $2$),
will contribute to the prey repopulation of cell $i$.
The third process represents the local depopulation of a cell
due to too greedy predators. It can be interpreted as the local extinction 
of the species or as the moving of them to neighboring occupied sites.
Spontaneous disappearance of a prey state ($\sigma_i$: $1 \to 0$)
or that of the predators alone ($\sigma_i$: $2 \to 1$)
is forbidden.
These assumptions are reasonable because the occurrence of these
processes is improbable.
The rate of the third process is chosen to be 1,
which sets the time scale.
As a consequence, $\lambda_a$ and $\lambda_b$ are 
dimensionless quantities.

The above dynamical rules are an extension of the 
contact process model (CP)~\cite{contact} introduced as a description of
epidemic spreading.
The CP is a 2-state model, $\sigma_i = 0, 1$;
the status 0 and 1 represent respectively
the healthy and the infected
individuals. The CP dynamical rules are
\begin{itemize}
\item[i)]
$0 \to 1$ at rate $\lambda (n_{i,1})/4$,
\item[ii)]
$1 \to 0$ at rate 1 ~.
\end{itemize}
An epidemic survives for $\lambda>\lambda_{CP}^*$ 
= $1.6488(1)$~\cite{grassberger} and disappears for $\lambda<\lambda_{CP}^*$.
The transition towards this absorbing state is of
second order and belongs to the directed percolation (DP)
universality class~\cite{DP}.

Our model differs from most of the lattice models previously
investigated~\cite{{tome},{boccara},{provata},{tainaka}} 
by the fact that on each site, each species may be represented
by several individuals rather than just one.
In the previously investigated models the spreading rate of the preys
is simply proportional to $n_{i,1}$.
Under this assumption, our model reduces essentially 
to the ST model, in which the control parameters are defined as
$c=(1+\lambda_a+\lambda_b)^{-1}$ and $p=c(\lambda_b-\lambda_a)/2 $.

It is worth discussing first the behavior of our model 
in two limiting cases.
In the $\lambda_a \to \infty$ limit the proportion of empty cells
is negligible since the empty cells are reoccupied by preys
instantly after their extinction. Hence, the lattice is completely
covered by preys and the $\sigma=2$ sites behave as the infected species
in the CP. Namely, when decreasing $\lambda_b$
the predator density is decreasing continuously and vanishes
at the CP critical value $\lambda_b^*(\lambda_a=\infty)$ 
= $\lambda_{CP}^*$.
One can think of the $\lambda_b \to \infty$ limit in similar terms.
In this case, the proportion of the prey cells ($\sigma=1$) 
should be negligible since the high productivity of the predators,
while the prey-predator cells should behave as the infected species in the CP.
This is indeed the case if $\lambda_a > \lambda_{CP}^*$,
but when $\lambda_a$ gets smaller than $\lambda_{CP}^*$,
the prey density increases again instead of being zero,
as we shall see later.

\section{Mean-field analysis.}
\label{sec:mean-field}

Although apparently simple, there is no way to solve analytically 
the model defined above. However, analytic solutions can be obtained
by making some approximations. The simplest one is the one-point 
mean-field approximation in which all spatial fluctuations are neglected. 
Thus, the system is characterized by the densities of prey, $a$,
and predator, $b$, sites
\be
a={1\over L^2} \sum_i ( \delta_{\sigma_i, 1} + \delta_{\sigma_i, 2} ) ~,~~
b={1\over L^2} \sum_i \delta_{\sigma_i, 2} ~,
\ee
which values satisfy the $0\le b \le a \le 1$ conditions by definition.
In terms of these densities, 
the mean-field dynamical equations read:
\be
\frac{da}{dt} = \lambda_a a(1-a)-b
\label{mfa}
\ee 
and 
\be
\frac{db}{dt} = \lambda_b b(a-b)-b
\label{mfb}
\ee 
Note, that for $b=0$ ($a=b=0$) initial condition the predator (and prey)
densities remains 0.

The (\ref{mfa},\ref{mfb}) equations clearly differ from the usual L-V
ones. The main difference lies in the interaction terms as, 
although a larger prey density increases the predator growth rate,
the rate of the predated preys only depends on the predator
density. This is a simple consequence of the fact 
that there are no pure predator sites without preys in this model.
Thinking of a real prey-predator system it makes sense, 
as a predator has to consume a certain amount of preys in a given time
to survive, independently of the number of preys around it.
The $(1-a)$ term in the first equation plays the role of a simple 
Verhlaust factor which assures an upper limit for the prey density 
$(a\le 1)$, and similarly the $(a-b)$ term in the second equation 
do not let the density of predators exceed that of the preys.

\begin{figure}[]
\centerline{
        \epsfxsize=8cm
        \epsfbox{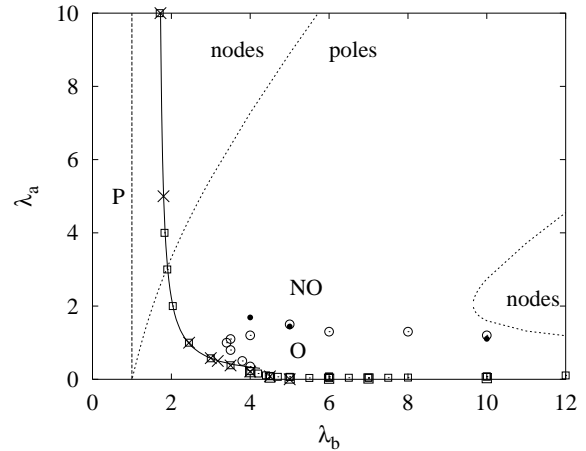}
           }
\figspace
\caption{Mean-field prediction for the
boundary (dashed line) between the prey (P) and the coexistence 
phase (O and NO).
The dotted lines are the boundaries between the pole and 
node type of stationary state regions.
The MC results are also depicted for comparison (see Fig.~\ref{fig:phase}
for the details).
}
\label{fig:phase_mf}
\end{figure}

The stationary states are obtained by setting the left hand sides of 
Eqs.~(\ref{mfa}, \ref{mfb}) to zero.
Contrary to the simplest L-V equations, 
qualitatively different stationary states are obtained
varying the parameters, $\lambda_a$ and $\lambda_b$,
as illustrated on Fig.~\ref{fig:phase_mf}.

For $0 \le \lambda_b \le 1$ and $\lambda_a>0$, 
the stationary state is a pure prey absorbing state $a^s=1, b^s=0$.
For $\lambda_a=0$ the stationary state 
is also a prey state, $b^s=0$, however, the value of $a^s$ 
depends upon the initial state.

In the rest of the plane $(\lambda_a, \lambda_b)$, the stationary 
solution is
\be
a^s=\frac{(\lambda_a -1) + \sqrt{(\lambda_a-1)^2 + 4 \lambda_a 
/\lambda_b}}{2 \lambda_a}
\ee
and 
\be
b^s=a^s-\frac{1}{\lambda_b} ~,
\ee
which describes a coexistence of preys and predators (coexistence phase).

For $\lambda_b\gg1$ the $a$ and $b$ densities are approximately the same
\be
a^s = b^s + O(\frac{1}{\lambda_b}) = \left\{ \ba{ll} 
\displaystyle
 1 - \frac{1}{\lambda_a} + O(\frac{1}{\lambda_b}) 
 & \quad \mbox{ for } \lambda_a > 1 \\
\displaystyle
 O(\frac{1}{\sqrt{\lambda_b}})  
 & \quad \mbox{ for } \lambda_a = 1 \\
\displaystyle
 O(\frac{1}{\lambda_b})  
 & \quad \mbox{ for } \lambda_a < 1 \\
\ea \right.
\ee
and as a function of $\lambda_a$ they show a mean field CP behavior
as it is expected from the argument given in Sec.~\ref{sec:model}.

In the $\lambda_a\gg1$ limit (and for $\lambda_b>1$)
the system is ''full of preys'', namely
\be
a^s = 1 - \frac{1}{\lambda_a}(1-\frac{1}{\lambda_b})
 + O(\frac{1}{\lambda_a^2})
\label{aapp}
\ee
and the predator density reads
\be
b^s = (1-\frac{1}{\lambda_a})(1-\frac{1}{\lambda_b}) 
+ O(\frac{1}{\lambda_a^2})
\label{bapp}
\ee
and, as expected, its $\lambda_b$ dependence agrees with the prediction
of the mean-field approximation for CP.
This approximation predicts a second order phase transition
along the whole $\lambda_b=1$ line, as in the $\lambda_b \to 1$ limit 
$a$ and $b$ approach linearly the values $1$ and $0$ respectively
\bea
a^s &=& 1 - \frac{\lambda_b-1}{\lambda_a+1} + O((\lambda_b-1)^2)\cr
b^s &=& \lambda_a \frac{\lambda_b-1}{\lambda_a+1} + O((\lambda_b-1)^2) ~.
\eea

The behavior of the densities is rather surprising
at the $\lambda_a=0$ boundary of the coexistence phase.
For $0<\lambda_a \ll 1$ and for $\lambda_b>1$
\be
a^s = \frac{1}{\lambda_b} + \lambda_a\left(\frac{1}{\lambda_b}
+ \frac{1}{\lambda_b^2}\right) + O(\lambda_a^2) ~,
\label{a:la<<1}
\ee
while the stationary solution, $a^s$, for $\lambda_a = 0$ 
depends on the initial state.
Thus the mean field approximation predicts a discontinuity
of the prey density along this boundary. 
However, the density $b^s = a^s - \lambda_b^{-1}$ 
is proportional to $\lambda_a$ and continuous in $\lambda_a = 0$.

Important quantities are the fluctuations of
the prey and the predator densities (mean square deviations),
which are normalized to be size independent for large systems
\be
\chi_\rho = L^2 \langle (\rho-\langle \rho \rangle)^2 \rangle, 
  \mbox{  with  } \rho = a \mbox{  or  } b ~,
\label{defchi}
\ee
and $\langle \rangle$ means the time average in the stationary state. 
For $\lambda_a, \lambda_b\gg1$ the majority of the sites are in state
$2$, with a few holes in it,
hence one can suppose that the holes are independent.
Consequently, the number of the holes follows a Poisson distribution,
from which the average hole number equals to the mean square deviation.
There are $L^2(1-a)$ holes made of sites in the state $\sigma_i = 0$
and $L^2(1-b)$ holes made of sites in the states $\sigma_i = 0$ or $1$.
Thus
\be
\chi_a \approx 1-a^s \mbox{~~and~~} \chi_b \approx 1-b^s ~,
\label{appchi}
\ee
which is in good agreement with the simulations in a region
(non-oscillatory part) of the coexistence phase (see Fig.~\ref{fig:chi}).

The stability of the stationary state can be analyzed by
linear stability. One has to investigate
the eigenvalues, $\epsilon_{1,2}$, of
the Jacobian matrix related to the mean field equations 
(\ref{mfa}, \ref{mfb}) at the stationary densities
\be
\left.
\left(\ba{cc}
\partial_a \dot a & \partial_b \dot a \\
\partial_a \dot b & \partial_b \dot b \\
\ea \right) \right|_s=
\left(\ba{cc}
\lambda_a(1-2a^s) & -1 \\
\lambda_b a^s -1 & 1 - \lambda_b a^s  
\ea \right) ~.
\ee

\begin{figure}[]
\centerline{
        \epsfxsize=6cm
        \epsfbox{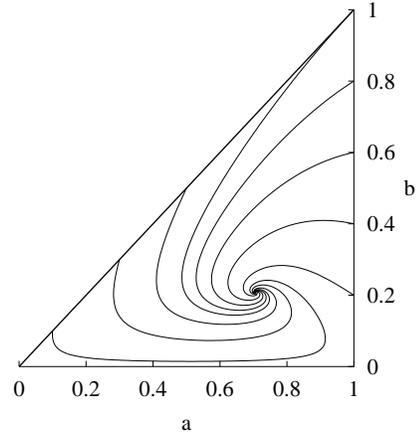}
           }
\figspace
\caption{Pole type of approach of the stationary solution 
in mean-field approximation for $\lambda_a=1$ and $\lambda_b=2$, 
starting the system from different initial conditions . 
Note, that the $0\le b\le a\le 1$ conditions are always satisfied.
}
\label{fig:flow}
\end{figure}

It turns out that
the real parts of the eigenvalues are always negative, 
assuring the stability of the solutions.
This mean field approximation
do not predict limit cycles, which would correspond to having
an eigenvalue, $\epsilon$, with a zero real part. 
However, in some part of the coexistence phase the imaginary part 
is nonzero, so the stationary solution is approached in spirals
(poles), instead of straight lines (nodes) 
(see Fig.~\ref{fig:phase_mf} and \ref{fig:flow}),
as it was also observed in the ST model~\cite{tome}. 
Note, that an unexpected node region appears for $\lambda_b>10$.
One can consider the presence of poles as a hint for the 
appearance of oscillations beyond the mean-field approximation.
Notice, that in this pole case, the damped oscillations are strong along the 
$\lambda_b$ axes (i.e. for $\lambda_a \ll 1$).
The strength of them can be characterized 
by the ratio of the imaginary and real part of the eigenvalues,
which has a singularity in the $\lambda_a \to 0$ limit.
Using (\ref{a:la<<1}), we obtain
\be
\left|\frac{\Im(\epsilon)}{\Re(\epsilon)}\right| =
4 \lambda_a^{-1/2} \sqrt{\lambda_b^2 - \lambda_b}
+ O (\lambda_a^{1/2}) 
\ee
for $\lambda_a \ll 1$ and $\lambda_b>1$.
In this limit one can derive an expression also for the frequency,
$\omega$, of the damped oscillations
\be
\omega = \left| \Im(\epsilon)  \right| =
2 \lambda_a^{1/2} \sqrt{1-\frac{1}{\lambda_b}}
+ O (\lambda_a^{3/2}) ~.
\ee

The mean field results can be interpreted in the following way.
The approximation predicts
two distinct phases: the pure prey phase and the coexistence one.
It also gives some hints for a possible presence of oscillations 
in some parts of the coexistence phase.
The phase boundaries of the two phases are described by two lines:
the $\lambda_b=1$ and the $\lambda_a=0$.
Several quantities show a power-law behavior close to these boundaries,
like $b$ and $1-a$ at the $\lambda_b=1$ boundary,
and $b$, $\omega$ and the strength of the damped oscillations at the
$\lambda_a=0$ boundary.
This implies that the transitions are of second order,
and the predator density, $b$, seems to be a good candidate for 
the order parameter.
The order parameter goes to zero at the phase boundaries
as $b \sim (\lambda_b-1)^\beta$ and $b \sim \lambda_a^\beta$
with a mean-field exponent $\beta=1$.

We performed also a pair approximation,
in which the nearest neighbor correlations are also considered as parameters.
It turns out that the results differ only quantitatively 
from that of the one point approximation.
Contrary to~\cite{tome}, our system does not show 
limit cycle behavior on the pair approximation level either.

\section{Monte-Carlo simulations.}
\label{sec:mc}

On general grounds, one expects that the fluctuations will play an important 
role in low dimensions. Our model is supposed to describe a two dimensional 
world and accordingly, we have performed extensive Monte-Carlo (MC)
simulations for
systems of sizes $L \times L$, $L$ varying between 100 and 1000. We used
periodic boundary conditions. 

\begin{figure}[]
\centerline{
        \epsfxsize=8cm
        \epsfbox{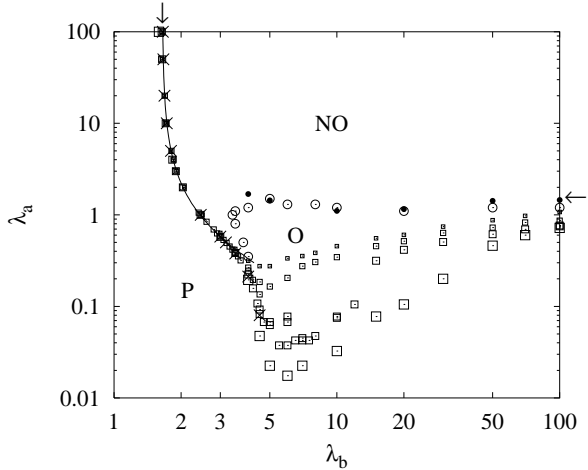}
           }
\figspace
\caption{Stationary state phase diagram as obtained by simulations.
The squares ($\Box$) indicates the transition to the prey absorbing state (P)
for different system sizes ($L=100$, 200, 500 and 1000),
and the arrows points to the $\lambda_{CP}^*$ value.
On all figures larger symbols correspond to larger systems.
The boundary between the oscillatory (O) and the non-oscillatory (NO)
region of the coexistence phase is determined based on Fourier 
analysis ($\circ$) and on the crossover in $\chi_a$ ($\bullet$).
For the DP type transition between P and NO,
the fitted values of $\lambda_b^*(\lambda_a)$ ($\times$)
and the approximation described in Sec.~\ref{sec:discuss} (solid line) 
is also depicted.
}
\label{fig:phase}
\end{figure}

Although our model is formulated as a continuous-time process,
an equivalent (at least for not very short times) 
discrete time formulation is more suitable for numerical simulations. 
In one elementary time step one lattice site is 
chosen randomly and its state evolves according to the rules 
defined is Sec.~\ref{sec:model} using rescaled rates (all less than 1)
as transition probabilities. One MC step is defined 
as the time needed such that all the sites have been, 
on the average, visited once. 
In this paper we always use the original time units defined by the model, 
which can be obtained simply by rescaling the time measured in MC steps. 

For sufficiently large system, the stationary state does not 
depend on the initial conditions.
Usually we filled up the lattice completely with preys as an initial state
and put a few predators on it.
To obtain the stationary phase diagram and the stationary values 
of the quantities of interest
the number of MC steps performed varied in the range $10^6$ 
to $10^5$ for systems of linear size $L=200$ to 1000 respectively.

\begin{figure}[]
\centerline{
        \epsfxsize=8cm
        \epsfbox{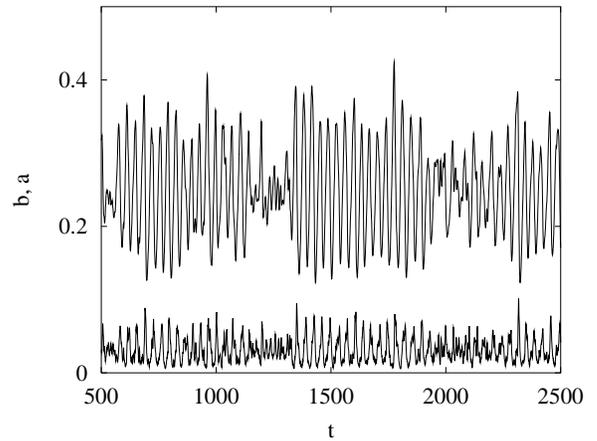}
           }
\figspace
\caption{Typical behavior of the prey, $a$, and the predator, $b$ ($b\le a$),
densities in the oscillatory region of the stationary state
($\lambda_a=0.8$, $\lambda_b=100$, $L=1000$). }
\label{fig:dens_time}
\end{figure}

\begin{figure}[]
\centerline{
        \epsfxsize=8cm
        \epsfbox{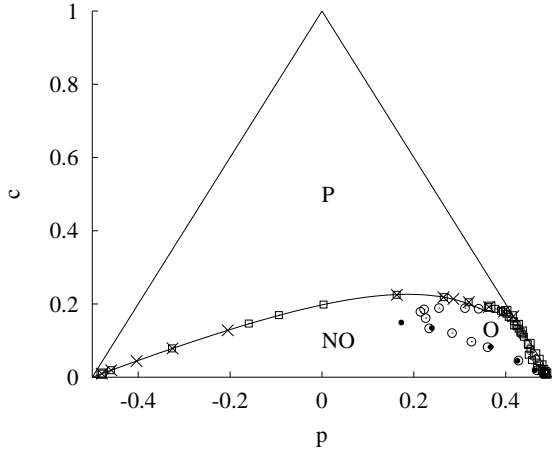}
           }
\figspace
\caption{The same as on Fig.~\ref{fig:phase} but as a function 
of the variables used in the ST model. The triangle
represents the available part of the phase space.
The location of the
oscillatory (O) and the non-oscillatory (NO) regions are quite 
different from that of the ST model.}
\label{fig:phase_t}
\end{figure}

The corresponding phase diagram is depicted on Fig.~\ref{fig:phase}. 
Two different phases are present 
as a function of the two control parameters $\lambda_a$ and $\lambda_b$:
a pure prey phase, a prey and predator coexistence phase 
with an oscillatory and a non-oscillatory region.
In the oscillatory region, oscillations with a well defined frequency
were observed in the prey and the predator densities 
(see Fig.~\ref{fig:dens_time}).
Although theoretically possible,
we never observed an empty lattice absorbing state.
The reason for that is simply that even one surviving
prey fill up the system with preys in the absence of predators.
As Fig.~\ref{fig:phase_t} shows, the locations of the different regions 
of the phase space differ essentially from those obtained for the ST model.

The phase boundaries of the prey phase (see Fig.~\ref{fig:phase})
were obtained in the following way.
Simulations were started at parameter values
for which the coexistence is maintained practically forever
(up to the maximal number of MC steps investigated),   
and we decreased one of the parameter values by $\Delta \lambda$.
If the predators were still alive after a given time, $\Delta t$,
we decreased the parameter further.
The extinction of the predators defines the phase boundary.
$\Delta \lambda$ was chosen to be in the range 0.005 to 0.04, 
while $\Delta t = 3\times 10^4$ MC steps.
The result was very similar with $\Delta t = 10^4$ 
and $5\times 10^4$ MC steps.
The definition of the boundary between the oscillatory and the 
non-oscillatory region of the coexistence phase will be described later.

On Fig.~\ref{fig:phase}, the boundary of the prey phase
is displayed for different system sizes ($L=100,..,1000$).
Apparently, in the $\lambda_a>\lambda_b$ regime 
the size dependence is negligible,
but relevant for $\lambda_a<\lambda_b$.
Note that this strong size dependence of the boundary
coincides with the presence of oscillations.

\begin{figure}[]
\centerline{
        \epsfxsize=8cm
        \epsfbox{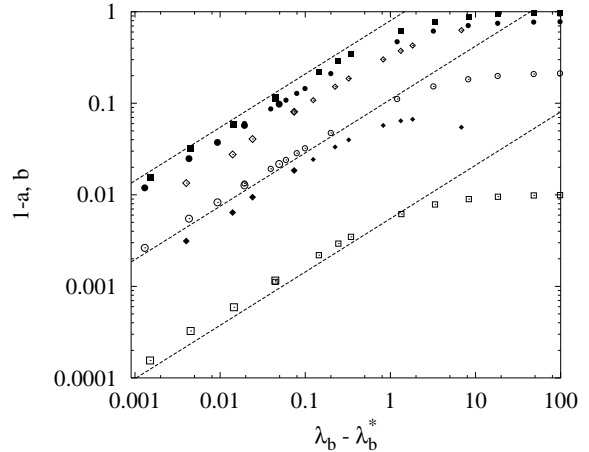}
           }
\figspace
\caption{Prey (open symbols) and 
the predator (filled symbols) densities
close to the second order phase transition line between
the prey phase and the non-oscillatory region of the coexistence phase.
$\lambda_a = 0.5(\Diamond)$, $5(\bigcirc)$ and $100(\Box)$ 
while the system sizes are $L=200$ and 500. 
The slope of the dashed lines is the
DP critical exponent $\beta \approx 0.583$.}
\label{fig:dens_dp}
\end{figure}

Decreasing $\lambda_b$ at any fixed value of $\lambda_a$, 
a second order phase transition takes place 
between the coexistence and the prey absorbing phases
along a transition line $\lambda_b^*(\lambda_a)$.
As for the mean field case, 
the predator density is considered to be the order parameter.
As $\lambda_b \to \lambda_b^*(\lambda_a)$, the order parameter, $b$,
and $1-a$ go to zero as
\be
b \sim 1-a \sim (\lambda_b - \lambda_b^*(\lambda_a))^{\beta_1} ~.
\label{dp_op}
\ee 
As seen on Fig.~\ref{fig:phase}, the values of $\lambda_b^*(\lambda_a)$
obtained by fitting the data with Eq.~\ref{dp_op} are in very good agreement
with the phase boundary obtained previously.
Fitting the data leads $\beta_1 \approx 0.58(1)$
(with satisfactory precision for $\lambda_a > 0.3$; see 
Fig.~\ref{fig:dens_dp}). 

In the same limit the fluctuations of the predator
density also follow a power law behavior,
$\chi_b \sim (\lambda_b - \lambda_b^*(\lambda_a))^{\gamma_1}$.
The exponent has been determined to a good precision as
$\gamma_1\approx 0.35(3)$ for several values of $\lambda_a$ between 1 and 50.
The same behavior has been obtained (only for $\lambda_a=1$ and 3)
for the prey fluctuations, $\chi_a$, with the exponent 
$\gamma_1\approx 0.35(5)$.
The critical behavior seems to be the same when the transition line is crossed
while decreasing $\lambda_a$ at fixed values of $\lambda_b$.
The two exponents, $\beta_1$ and $\gamma_1$, are compatible with those 
obtained for DP in $2+1$ dimension~\cite{jensen}.
Thus we conclude, that this absorbing state phase transition belongs to 
the DP universality class, as expected on general grounds~\cite{DPconj}.
Note that DP type phase transition in a similar
lattice prey-predator system has already been observed
in 1 dimension~\cite{lipowska}.

\begin{figure}[]
\centerline{
        \epsfxsize=8cm
        \epsfbox{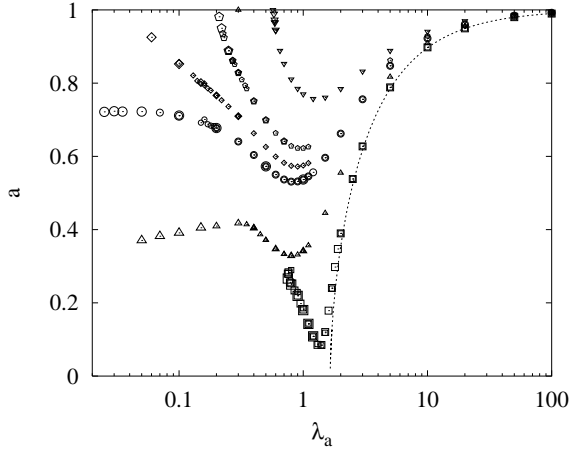}
           }
\figspace
\caption{Prey density, $a$, as a function of $\lambda_a$ 
for different values of $\lambda_b = 3 (\bigtriangledown)$, $4$ (pentagon), 
$4.5 (\Diamond)$, $5 (\bigcirc)$, $10 (\triangle)$, $100 (\Box)$ and
system sizes $L=200$, 500, 1000. The dashed line is the density
given by the CP.}
\label{fig:dens_a}
\end{figure}

\begin{figure}[]
\centerline{
        \epsfxsize=8cm
        \epsfbox{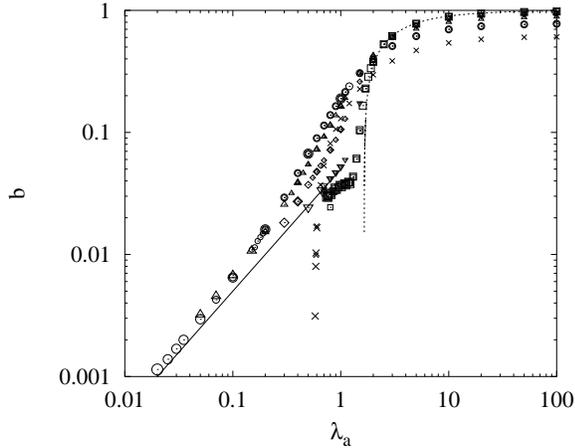}
           }
\figspace
\caption{Predator density, $b$, as a function of $\lambda_a$ 
for different values of $\lambda_b = 3(\times)$, $5 (\bigcirc)$, 
$10 (\triangle)$, $20(\Diamond)$, $50(\bigtriangledown)$, $100 (\Box)$ and
system sizes $L=200$, 500, 1000 and 2000 only for $\lambda_b=5$. 
The $\lambda_a \to 0$ behavior
is close to a power law with an exponent 1 (solid line),
while the dashed line is the density given by the CP.}
\label{fig:dens_b}
\end{figure}

For $\lambda_a \to 0$, the transition line, $\lambda_b^*(\lambda_a)$,
ends in a special point, 
($\lambda_a=0,\lambda_b^T\approx 5.0(3)$),
where all the three phases meet.
For $\lambda_b>\lambda_b^T$, 
the MC results for the finite size dependent phase boundary suggest us
that the transition happens at $\lambda_a^* = 0$ in the $L\to\infty$ limit.
Approaching this transition line, $\lambda_a \to 0$, the prey density, $a$, 
does not go to 1 but to a finite value depending on $\lambda_b$ 
(see Fig.~\ref{fig:dens_a}). 
However, according to the results depicted on Fig.~\ref{fig:dens_b},
the predator density, $b$, always goes to zero in this limit 
as a power of $\lambda_a$ with an exponent $\beta_2 \approx 1$.
Surprisingly, this second order phase transition to the prey
absorbing phase does not belong to the DP universality class.
The presence of power law behavior, however, 
confirms that for an infinite system
the transition occurs at $\lambda_a^* = 0$ for $\lambda_b > \lambda_b^T$.
This means that for this range of $\lambda_b$, and
for any arbitrary small $\lambda_a$, the coexistence of the species 
is possible providing that the system is large enough.

For $\lambda_b>\lambda_b^T$ 
the fluctuations of the two densities, $\chi_a$ and $\chi_b$, 
behaves similarly.
For a given $\lambda_b$, there is a clear crossover at
$\lambda_a^O(\lambda_b)$ from a mean field like behavior
to a regime where the correlations are more important.
For $\lambda_a > \lambda_a^O(\lambda_b)$
the behavior of $\chi_a$ and $\chi_b$ agrees with that predicted by mean-field
theory, reflecting the fact that in this range of $\lambda_a$
the dominant behavior comes from the noise.
As the $\lambda_a < \lambda_a^O(\lambda_b)$ condition
coincides with the presence of oscillations, the crossover point, 
$\lambda_a^O(\lambda_b)$, is taken as the definition of the border 
between the oscillatory and the non-oscillatory region.

\begin{figure}[htb]
\centerline{
        \epsfxsize=8cm
        \epsfbox{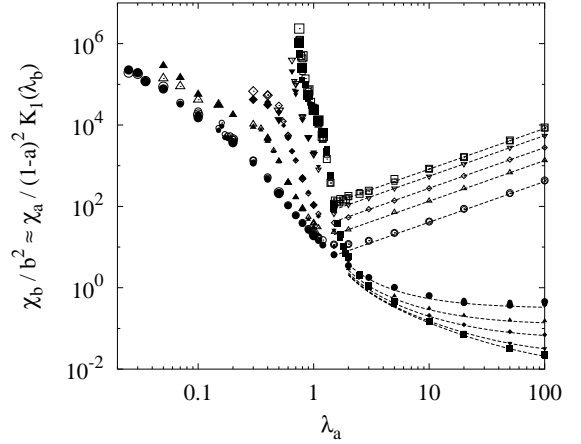}
           }
\figspace
\caption{Normalized fluctuations of the prey (open symbols) and 
predator (filled symbols) densities,
$K_1(\lambda_b)\chi_a/(1-a)^2$ and $\chi_b/b^2$, which collapse in the
oscillatory region. The parameters are $\lambda_b = 5 (\bigcirc)$, 
$10 (\triangle)$, $20(\Diamond)$, $50(\bigtriangledown)$, $100 (\Box)$ and
system sizes $L=200$, 500, 1000. The dashed lines correspond
to the mean-field solutions (\ref{appchi}).
}
\label{fig:chi}
\end{figure}

After a proper normalization, the relative fluctuations
collapse on a single curve for $\lambda_a<\lambda_a^O(\lambda_b)$
(see Fig.~\ref{fig:chi}). Namely,
\be
\frac{\chi_b}{b^2} \approx  K_1(\lambda_b) \frac{\chi_a}{(1-a)^2} ~,
\ee
where the numerical factor, $K_1(\lambda_b)$, depends only on $\lambda_b$.
However, the precision of the simulation results was not satisfactory 
enough to obtain the functional form of $K_1(\lambda_b)$ 
(and of the forthcoming $K_i(\lambda_b)$ for $i=2$ ,3 and 4 either).

The simulation showed that $\chi_\rho$ ($\rho=a$ or $b$) is size independent
as it was expected from its definition (\ref{defchi}).
As a consequence,
the deviation from the average density, $\sigma = \sqrt{\chi_\rho}/L$
\cite{tome}, 
is smaller for larger systems and evidently scales with $1/L$.
Certainly, this deviation increases with the intensity of the oscillations.
The above finite size behavior is in agreement with the results of earlier 
simulations which claimed that the oscillations in the global densities
disappear with increasing system size~\cite{boccara}.
Our simulations predict more pronounced oscillations
for smaller $\lambda_a$ and for larger $\lambda_b$.

The oscillations have to show up also in the correlation functions,
\bea 
C_a(i, \tau) &=& \langle (1 - \delta_{\sigma_j(t), 0})
	(1 - \delta_{\sigma_{j+i}(t+\tau), 0}) \rangle ~, \cr\cr
C_b(i, \tau) &=& \langle \delta_{\sigma_j(t), 2} 
	\delta_{\sigma_{j+i}(t+\tau), 2} \rangle ~,
\eea
where $j+i$ labels a lattice site distant of $i$
lattice spacing from the site $j$.
$C_\rho$ ($\rho=a$ or $b$) depends only on $i$ and $\tau$
due to the homogeneity of the system in space and time.
For $\tau=0$ the correlation function, $C_\rho(i)$,
obtained numerically could be fitted by an exponential
$C_\rho(i) \sim \exp(-i/\xi_\rho)$.
In the oscillatory region the correlation lengths of preys and predators
differ only through a $\lambda_b$ dependent factor, 
$\xi_a\approx K_2(\lambda_b)\xi_b$, 
and they turned out to be proportional to the 
fluctuations of the prey density, $\xi_a \approx \sqrt{2\chi_a}$.
It means that a more correlated system displays stronger oscillations.
The reason for that is simply that the dynamics of the different sites 
shows some synchronization within a correlation length,
which results in larger oscillations
(see Sec.~\ref{sec:discuss} for more details).

\begin{figure}[htb]
\centerline{
        \epsfxsize=8cm
        \epsfbox{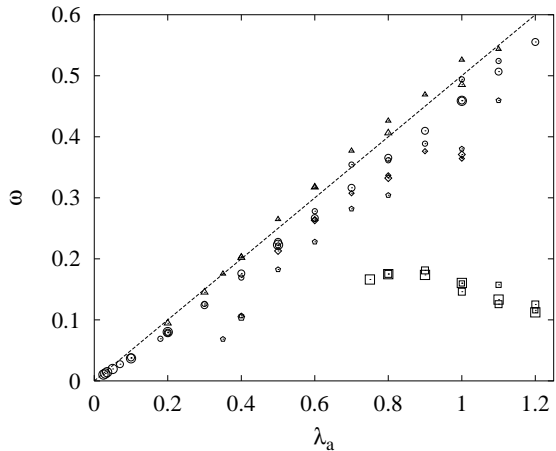}
           }
\figspace
\caption{Frequency of the oscillations as a function of $\lambda_a$
and for $\lambda_b=4$ (pentagon), $5 (\bigcirc)$, 
$10 (\triangle)$, $20(\Diamond)$, $100 (\Box)$ and
for system sizes $L=200, 500, 1000$. For a wide range of the parameters
the data are close to $\lambda_a/2$ (dashed line).}
\label{fig:freq}
\end{figure}

In order to determine the characteristic frequency, 
$\omega_\rho(\lambda_a, \lambda_b)$, and the amplitude, 
$A_\rho(\lambda_a, \lambda_b)$ ($\rho=a$ or $b$), of the oscillations,
we measured the Fourier spectrum of the time dependent densities
\be
S_{\rho}(\omega) = \lim _{T\to \infty} {1 \over T} 
 \left| \sum_{t=1}^T \rho(t) \exp(i\omega t) \right|^2 ~.
\label{defps}
\ee
The presence of oscillations is reflected as a peak
at nonzero frequency in the Fourier spectrum.
Extracting this peak from a background noise,
enable us to define $A_\rho$ and $\omega_\rho$ as the zeroth and the first 
momentum of this distribution. 
This analysis shows clearly 
that the frequency of the oscillations is independent
of the system size (see Fig.~\ref{fig:freq}),
and is the same for preys and predators. 
Moreover, for a wide range of the parameters 
in the oscillatory phase the frequency, $\omega=\omega_a=\omega_b$, 
is well approximated by $\lambda_a/2$.
This linear behavior differs from the mean field prediction.

\begin{figure}[htb]
\centerline{
        \epsfxsize=8cm
        \epsfbox{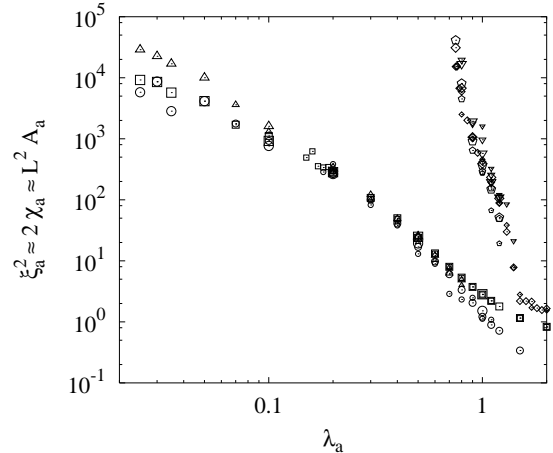}
           }
\figspace
\caption{Test of the relation between several characteristics of
the prey population (see Eq.~\ref{xca}), namely the correlation length,
$\xi_a (\triangle, \bigtriangledown)$, the fluctuations of the prey density,
$\chi_a (\Box, \Diamond)$, and the amplitude of the oscillations,
$A_a$ ($\bigcirc$, pentagon) for $\lambda_b=5$ and 100 respectively.
The sizes of the system are $L=200$, 500 and 1000.
}
\label{fig:scale}
\end{figure}

In the oscillatory region the oscillations are present
for arbitrary large systems, however, their amplitude decreases
with increasing system size, as $1/L^2$.
At this point it is important to emphasize that this fact does not
imply that only small oscillations are present in large systems.
Indeed, for a large system the amplitude of the oscillations 
can be made larger by decreasing $\lambda_a$.
On the other hand, when increasing $\lambda_a$ the amplitude goes to zero
as a power law which makes difficult to define a phase boundary 
for the oscillations in this way.
However, there is a simple scaling relation between the amplitude 
and the correlation length for the preys in the oscillatory region
\be
\xi_a^2 \approx 2 \chi_a \approx L^2 A_a ~,
\label{xca}
\ee
as it can be observed on Fig.~\ref{fig:scale}.
The analogous expression for the predators is slightly more complicated
\be
\xi_b^2 \left(\frac{b}{1-a}\right)^2 K_3(\lambda_b)\approx \chi_b 
\approx K_4(\lambda_a) L^2 A_b ~,
\label{xcb}
\ee
with appropriate $K_3(\lambda_b)$ and $K_4(\lambda_a)$ values.

Another quantity which characterizes the oscillations is
the time dependent local correlations,
$C_\rho(\tau) = C_\rho(i=0, \tau)$. 
A similar investigation was made in~\cite{provata} 
with time dependent correlations of the average local densities.
In the oscillatory region $C_\rho(\tau)$ displays 
damped, size independent oscillations.
More precisely, the time correlations are size independent
for any $L>L_c(\lambda_a, \lambda_b)$, while for any $L<L_c$
the system evolves to the prey absorbing state.
Clearly, this critical system size is proportional to the
correlation length, $L_c\sim\xi$.
The size independence of $C(\tau)$ is a simple consequence
of the fact that areas which are further than $\xi$ apart
are uncorrelated.
The investigation of the time dependent correlations, however, 
provides a rather ambiguous way to
define the boundary of the oscillatory region.
Indeed, one can observe local oscillations everywhere in the 
coexistence phase simply because, due to the cyclic dominance nature
of the model, each site has to evolve
in a loop ($\sigma=0 \to 1 \to 2 \to 0 \dots$).
Thus, according to the value of the damping factor,
it is somehow arbitrary to decide if the state is 
oscillatory or not.

It is worth noting, that at some particular values of $\lambda_b$
($\lambda_b=10$ or 20) and for small $\lambda_a$ values ($\lambda_a <0.2$ or
0.4 respectively),
where the correlation length is comparable to the system size ($L\sim 500$),
the system can evolve to a stripe like state.
In this state 3 stripes of size $L$, made of predator, prey and empty cells,
are drifting through the system.
However, for given $\lambda_a$ and $\lambda_b$ values, 
this behavior disappears when increasing the size of the system. 

The comparison of the MC results with the mean-field prediction
shows that the later gives a qualitatively correct description of 
the phase diagram (see Fig.~\ref{fig:phase_mf}),
as well as of the discontinuity in the prey density, $a$, 
along the $\lambda_a=0$ boundary.

\section{Discussion}
\label{sec:discuss}

A qualitative understanding of the phase diagram is possible.
If the birth rates are much larger than the death rate 
($\lambda_a \gg 1$ and $\lambda_b \gg 1$) 
the system is full of preys and predators ($a \approx b \approx 1$),
while for small values of $\lambda_b$ the system evidently reaches
the pure prey absorbing state.

As already discussed in Sec.~\ref{sec:model},
in the $\lambda_a \to \infty$ the system is full of preys ($a \to 1$)
and the predators behave like the infected species in the CP.
It means that they could survive only for $\lambda_b > \lambda_{CP}^*$, 
where a DP like second order transition occurs.
This is in agreement with the mean field results
and with the simulation for $\lambda_a=100$ (see Fig.~\ref{fig:dens_dp}).

\begin{figure}[htb]
\centerline{
        \epsfxsize=6cm
        \epsfbox{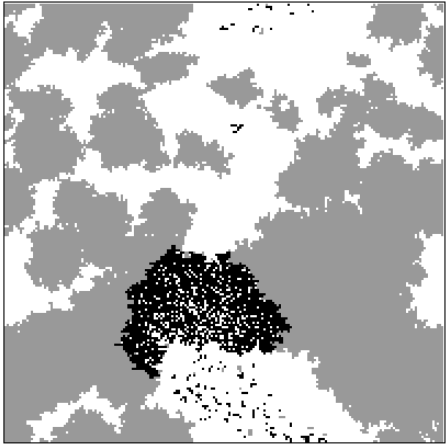}
            }
\figspace
\caption{Typical stationary state configuration 
of preys (grey) and predators (black)
on a 200 $\times$ 200 lattice at $\lambda_a=0.9$ and $\lambda_b=100$.
The white parts represent the empty sites. The picture shows
the beginning of the invasion of the pure prey territory by predators,
which were screened by empty sites before.}
\label{fig:conf100}
\end{figure}

One can also derive an approximate formula for the position
of the phase boundary between the non-oscillatory 
and the prey phase $\lambda_b^*(\lambda_a)$.
For $\lambda_a\gg 1$, 
the system is almost full of preys ($a \approx 1$)
and, in some sense, the dynamics of the predators is close 
to that of the CP.
The predators die at rate 1 and spread at rate $\lambda_b$, but
they cannot enter into the empty sites. 
One can introduce an effective $\tilde{\lambda_b}$
and describe the process as a CP, 
namely, the predators can enter any neighboring site at this rate.
As the number of empty sites is proportional to leading order to
$1/\lambda_a$, the effective parameter should be
$\tilde{\lambda_b} = \lambda_b - c/\lambda_a$,
where $c$ is a fitting parameter.
As this CP displays a phase transition at 
$\tilde{\lambda_b} = \lambda_{CP}^*$, in terms of the original 
parameter the transition occurs at 
$\lambda_b^*(\lambda_a) = \lambda_{CP}^* + c/\lambda_a$.
This conjecture is in excellent agreement with the simulation data
for $\lambda_a>0.5$ with $c=1.28(3)$ (see Fig.~\ref{fig:phase}).

\begin{figure}[htb]
\centerline{
	\epsfxsize=6cm
       	\epsfbox{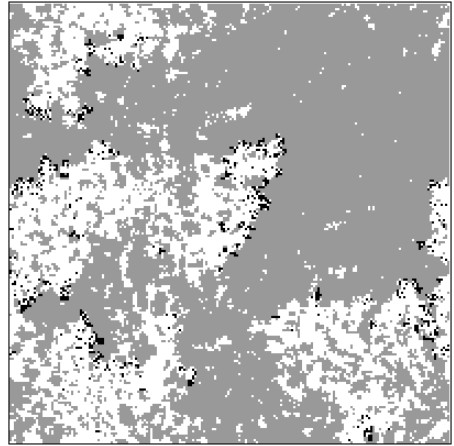}
           }
\figspace
\caption{The same as Fig.~\ref{fig:conf100} but for 
$\lambda_a=0.2$ and $\lambda_b=5$. Note, that the predators invade only 
the fully dense prey areas on both figures.}
\label{fig:conf5}
\end{figure}

For $\lambda_b \gg 1$ the new prey sites are usually immediately
occupied by predators as well. However, with a small but finite probability,
a predator site can disappear before the predators spread to the new 
born prey site, and in this way, a prey site can be left alone
and grow (similarly to the Eden model~\cite{eden}).
This rare event is negligible when the predator density is large enough
and a prey island cannot grow for long periods of time.
Practically this is the case for $\lambda_a > \lambda_{CP}^*$.
In this case, the number of prey sites is negligible small,
and the predator sites behave as the infected species in the CP.
One can see on the Fig.~\ref{fig:dens_a} and \ref{fig:dens_b},
that for $\lambda_b=100$ the two densities ($a \approx b$)
are equal to that of the CP if $\lambda_a > \lambda_{CP}^*$. 
However, in the vicinity of $\lambda_{CP}^*$ the densities 
are low, which allows an isolated prey island to grow
for a long time. If $\lambda_a < \lambda_{CP}^*$ the predator islands
are shrinking and, if $\lambda_a$ and $L$ are not too small, 
they can survive until a growing prey island reaches one of them.
At this moment, the predators invade very quickly the prey territory
and increase their population size (see Fig.~\ref{fig:conf100}).
These new predator sites start to die out leaving a few prey sites alone,
and the whole procedure starts again.
This mechanism insures the survival of the predators
much bellow the CP critical density and results in oscillations in
the population sizes.

For $\lambda_b > \lambda_b^T$, but not too large, the qualitative picture is 
slightly different. As one can observe on Fig.~\ref{fig:conf5}, groups
of predators are wandering through the system towards
prey-dense areas. 
If two fronts of predators meet they usually stop moving 
and the local population of predators starts shrinking.
The oscillations are maintained in a somewhat similar way than 
for the $\lambda_b \gg 1$ case:
these predators can only survive if the preys become dense around them.
This is more probable for larger values of $\lambda_a$, and
it is also clear that the predators have a better chance to 
survive in larger systems.

According to the above statements,
the key point in the underlying mechanism of oscillations
is the existence of blocked predator islands which are located 
and trapped in sparse prey areas.
Indeed, blocked predators in sparse prey areas result in growing 
prey populations; however, the resulting dense prey population 
allows predators to move and predate again. This mechanism 
drives back the system to the beginning of the loop.
Clearly, predators can only be trapped in sparse prey areas
if $\lambda_a$ is smaller or of the order of the death rate, 1.
This explains the location of the oscillatory region.
Note, that the above argument is based on the spatial
nature of the system, suggesting that the spatially extended 
character is fundamental for the existence of such prey-predator type 
of oscillations.

This mechanism also provides a qualitative understanding of
the key properties of the system.
The trapped predators can invade the prey area only
when the preys are dense enough again, which takes a time 
proportional to $1/\lambda_a$, and leads to $\omega\sim\lambda_a$.
According to the simulations the correlation length, $\xi$, 
increases with decreasing $\lambda_a$.
Indeed, as $\lambda_a$ decreases, 
the trapped predators have to wait longer
to escape, hence fewer groups of predators survive.
This increases the distance between the groups,
resulting in larger prey islands,
which average size is proportional to $\xi_a$. 

When the correlation length is of the order of the system size,
there are islands of preys of typical size L,
extruding the predators out of the system.
Hence, the condition $\xi_a\sim L$ characterizes the phase boundary 
between the oscillatory
and the prey phase. On the other hand,
a correlation length of order one ($\xi_a\sim 1$),
means that the noise dominates the system.
Thus, $\xi_a\sim 1$ characterizes the boundary between the oscillatory 
and the non-oscillatory region of the coexistence phase.

As shown by the study of the time dependent correlations,
domains separated by a distance larger than $\xi_a$ oscillate asynchronously
around a constant value with the same frequency, 
$\omega(\lambda_a, \lambda_b)$.
According to this picture, one can derive a more quantitative 
description for the oscillatory region. 
Let us assume that for $1\ll \xi_a \ll L$, the global densities 
of each species can be written as the sum of local coarse-grained 
densities at a typical length scale $\xi_a$.
Moreover, we assume that all these local densities oscillate with
the same frequency but a different phase, $\alpha_l$.
In general, the amplitude of the local oscillations should
depend on the parameters $\lambda_a$ and $\lambda_b$.
However, as one can observe on Fig.~\ref{fig:conf100} and \ref{fig:conf5},
the predators can only enter an almost fully dense prey area
and the predator fronts leave an almost empty field behind them.
Hence, as suggested by the numerical simulations, everywhere
in the oscillatory region, the local amplitude for the prey 
density can be considered as a constant, $d$. Thus
\be
a(t) = a^s + d \left(\frac{\xi}{L}\right)^2
 \sum_{l=1}^{\left(\frac{L}{\xi}\right)^2} 
 \sin(\omega t + \alpha_l) ~.
\ee
Supposing that the $\alpha_l$ values change much more slowly
than $\omega$, $a(t)$ shows a simple $\sin$ behavior
for long periods of time (see Fig.~\ref{fig:dens_time}).
Thus, for $a(t)$ one can derive the value of 
the density fluctuations, $\chi_a$, and the amplitude of these oscillations, 
$A_a$, using (\ref{defchi}) and (\ref{defps}),
and take the average over all the possible $\alpha_l$ configuration
taken from a flat distribution.
This procedure reproduces the result of Eq.~(\ref{xca}) up to
a multiplicative factor in front of the correlation length.

\section{Conclusions}
\label{sec:conc}

We have studied a two dimensional prey-predator model, (size $L \times L$), 
which exhibits a rich stationary state phase diagram. A particular 
attention has been payed to the study of finite size effects, 
and we were able to draw clear cut conclusions concerning the 
behavior of the model both for $L$ finite as well as for the 
limit $L \to \infty$.

Three kinds of stationary states can be reached according to the 
values of the control parameters : a pure prey state, and two  
coexisting prey-predator ones with and without oscillations.
Two different kinds of second order transitions were found when 
going into the  prey absorbing phase. 
The transition between  oscillatory and non-oscillatory coexistence 
phase is, in fact, a cross-over between 
two asymptotic regimes characterized by a very small and a large 
correlation length respectively.
In the oscillatory regime, scaling relations were established 
between several physical quantities.

A qualitative explanation for the existence of such oscillatory  
regime is given, pointing out the crucial role of the spatial 
extension  of the system.
Indeed, the frequency of the oscillations is determined locally due
to the dynamics related to blocked predator islands in sparse prey areas.
Regions of linear size $\xi_a$ oscillate with the same frequency
but with different phases. This explains the decreasing amplitude 
of oscillations with increasing system size.
On the other hand, slowly changing phases result periodic oscillations
in the overall prey density for long periods of time. 
Moreover, for suitable choices of the control parameters one can have 
synchronized oscillations with finite amplitude 
across arbitrary large systems. Thus we think that our simple model could 
offer a qualitative explanation for the behavior of the 
lynx population problem described in Sec.~\ref{sec:intro}.

\acknowledgements

We thank Z. R\'acz and G. Szab\'o for helpful discussions.
This work has been partially supported by the Swiss National Foundation.


\end{multicols}
\end{document}